\documentclass[12pt]{article}
\usepackage{amsmath,amssymb,amsthm,amsxtra,bbm,overpic,bm,epsfig,subfigure}
\usepackage{mathrsfs}
\usepackage{enumitem}
\usepackage{graphicx}
\usepackage{xcolor}
\usepackage{comment}
\usepackage{epstopdf}
\usepackage{float}
\usepackage{cite}
\allowdisplaybreaks[2]
\usepackage[all]{xy}
\usepackage{slashed,stmaryrd,multirow}
 \numberwithin{equation}{section}

\def\thefootnote{\fnsymbol{footnote}}

\newcommand{\Rcal}{\mathcal{R}}

\newcommand{\diag}{\operatorname{diag}}

\usepackage{hyperref}
\definecolor{linkblue}{HTML}{0D387F}
\definecolor{citegreen}{HTML}{006B47}
\definecolor{urlpurple}{HTML}{6B1F8C}
\hypersetup{
	colorlinks=true,
	linkcolor=linkblue,
	citecolor=citegreen,
	urlcolor=urlpurple,
	filecolor=urlpurple,
	pdfborder={0 0 0}
}
\usepackage{slashed,stmaryrd,orcidlink}

\usepackage{lscape}%
\usepackage{array}
\usepackage{booktabs}%

\begin{document}
	
	\vspace{0.2cm}
	
	\begin{center}
		{\Large\bf Deciphering Matter Invariants via Renormalization Group Equations for Neutrino Oscillations}
	\end{center}
	
	\vspace{0.2cm}
	
	\begin{center}
		{\bf Xin Wang}~{\orcidlink{0000-0003-4292-460X}}~$^{1,~2}$~\footnote{E-mail: xin.wang@unipd.it},
		\quad
		{\bf Shun Zhou}~{\orcidlink{0000-0003-4572-9666}}~$^{3,~4}$~\footnote{E-mail: zhoush@ihep.ac.cn}
		\\
		\vspace{0.2cm}
		{\small
			$^{1}$Dipartimento di Fisica e Astronomia ``Galileo~Galilei'', \\
			Università degli Studi di Padova, Via Francesco~Marzolo 8, 35131 Padova, Italy \\
			$^{2}$INFN, Sezione di Padova, Via Francesco~Marzolo 8, 35131 Padova, Italy\\
			$^{3}$Institute of High Energy Physics, Chinese Academy of Sciences, Beijing 100049, China\\
			$^{4}$School of Physical Sciences, University of Chinese Academy of Sciences, Beijing 100049, China}
	\end{center}

	\vspace{0.5cm}
	
		\begin{abstract}

			We utilize renormalization group equations (RGEs) for neutrino oscillations in matter to decipher the structure of exact matter invariants. 
			By combining the RGEs with the $S^{}_3$ permutation covariance under relabeling of the neutrino mass eigenstates, we recast all five algebraically independent matter invariants in the three-flavor framework as exact first integrals. Treating the matter potential as a matter spurion and imposing the cancellation conditions for the $1/\widetilde\Delta_{ij}^{}$ poles in the RGEs, we further prove that the three-flavor framework contains only two independent monomial invariants, which can be related to the Naumov and Toshev relations. We then extend the analysis to the four-flavor framework, where we uncover a complete set of eleven algebraically independent matter invariants and prove that the rank-two electron--sterile spurion obstructs the common pole cancellation required for any nontrivial multiplicative monomial invariant.
		\end{abstract}

	\def\thefootnote{\arabic{footnote}}
	\setcounter{footnote}{0}

    \newpage
    
	\section{Introduction}
	\label{sec:intro}

	Over the past few decades, neutrino oscillations have been extensively observed in a wide variety of experiments, firmly establishing that neutrinos are massive and there is significant flavor mixing in the lepton sector~\cite{Xing:2011zza,ParticleDataGroup:2026mpi,Xing:2020ijf}. Most recently, with only $59.1$ days of data, the Jiangmen Underground Neutrino Observatory (JUNO) simultaneously measured $\sin^2\theta_{12}^{}=0.3092\pm0.0087$ and $\Delta m_{21}^2 \equiv m^2_2 - m^2_1 =(7.50\pm0.12)\times10^{-5}~{\rm eV}^2$, corresponding to relative precisions of $2.8\%$ and $1.6\%$, respectively~\cite{JUNO:2025gmd}.\footnote{The first JUNO hint of $\Delta m_{31}^2 \equiv m^2_3 - m^2_1>0$ has recently been reported at the $2.3\sigma$ level~\cite{Wang:2026JUNO}.} As neutrino physics enters an era of precision measurements, matter effects in neutrino oscillations play an increasingly important role in the analysis and interpretation of experimental data.
	
	Matter effects~\cite{Wolfenstein:1977ue,Mikheyev:1985zog} arise from coherent forward scattering between propagating neutrinos and background particles, which modifies the propagation of neutrino mass eigenstates and thus the oscillation probabilities. In the standard three-flavor framework, after subtracting the flavor-universal neutral-current contribution, the effective Hamiltonian in the flavor basis is~\cite{Barger:1980tf,Kuo:1989qe}
	\begin{align}
	H(a)
	=
	\frac{1}{2E}
	\left[
	U
	\begin{pmatrix}
	m_1^2 & 0 & 0\\
	0 & m_2^2 & 0\\
	0 & 0 & m_3^2
	\end{pmatrix}
	U^\dagger
	+
	\begin{pmatrix}
	a & 0 & 0\\
	0 & 0 & 0\\
	0 & 0 & 0
	\end{pmatrix}
	\right]
	=\frac{1}{2E}
	V
	\begin{pmatrix}
	\widetilde m_1^2 & 0 & 0\\
	0 & \widetilde m_2^2 & 0\\
	0 & 0 & \widetilde m_3^2
	\end{pmatrix}
	V^\dagger
	\;,
	\label{eq:threeHamiltonian}
	\end{align}
	where $E$ denotes the neutrino energy, while $U$ and $m_i^{}$ (for $i = 1, 2, 3$) are the mixing matrix and neutrino masses in vacuum, respectively. The corresponding effective quantities in matter are denoted by $V$ and $\widetilde m_i^{}$ (for $i = 1, 2, 3$). The matter parameter is defined as $a\equiv 2\sqrt{2}G_{\rm F}^{}N_e^{}E$, where $G_{\rm F}^{}$ is the Fermi constant and $N_e^{}$ is the electron number density.

	Although the three-flavor Hamiltonian in matter can be diagonalized exactly~\cite{Zaglauer:1988gz,Krastev:1988yu,Xing:2000gg,Xing:2016ymg}, one may instead seek combinations of physical quantities that are independent of $a$. Such matter invariants yield exact relations that retain the same form in vacuum and matter. Two examples are particularly interesting. The first one is the Naumov relation~\cite{Naumov:1991ju,Krastev:1988yu,Harrison:1999df,Xing:2000ik}
	\begin{equation}
	\widetilde I_{\rm N}^{}
	\equiv
	\widetilde J\,
	\widetilde\Delta_{12}^{}
	\widetilde\Delta_{23}^{}
	\widetilde\Delta_{31}^{}
	=
	J\,
	\Delta_{12}^{}
	\Delta_{23}^{}
	\Delta_{31}^{}
	\equiv I_{\rm N}^{}
	\;,
	\label{eq:NaumovRelationIntro}
	\end{equation}
	where $\widetilde\Delta_{ij}^{}\equiv\widetilde m_i^2-\widetilde m_j^2$ and $\widetilde J\equiv{\rm Im}(V_{e1}^{}V_{\mu2}^{}V_{e2}^{*}V_{\mu1}^{*})$ denote the effective mass-squared differences and the Jarlskog invariant~\cite{Jarlskog:1985ht,Wu:1985ea} in matter, respectively, with $\Delta_{ij}^{}$ and $J$ being their vacuum counterparts. The second example is the electron-row invariant~\cite{Kimura:2002wd}
	\begin{equation}
	\widetilde I_e^{}
	\equiv
	|V_{e1}^{}|^2|V_{e2}^{}|^2|V_{e3}^{}|^2
	\widetilde\Delta_{12}^2
	\widetilde\Delta_{23}^2
	\widetilde\Delta_{31}^2
	=
	|U_{e1}^{}|^2|U_{e2}^{}|^2|U_{e3}^{}|^2
	\Delta_{12}^2
	\Delta_{23}^2
	\Delta_{31}^2
	\equiv I_e^{}
	\;,
	\label{eq:ElectronRowInvariantIntro}
	\end{equation}
	which, together with Eq.~(\ref{eq:NaumovRelationIntro}), leads to the Toshev relation~\cite{Toshev:1991ku}
    \begin{equation}
	\widetilde I_{\rm T}^{}
	\equiv
	\frac{\widetilde I_{\rm N}^{}}
	{(\widetilde I_e^{})^{1/2}}
	=
	\frac{\widetilde J}
	{|V_{e1}^{}||V_{e2}^{}||V_{e3}^{}|}
	=
	\frac{J}
	{|U_{e1}^{}||U_{e2}^{}||U_{e3}^{}|}
	=
	\frac{I_{\rm N}^{}}
	{(I_e^{})^{1/2}}
	\equiv I_{\rm T}^{}
	\;.
	\label{eq:ToshevRelationIntro}
	\end{equation}
	Hereafter, we use $(I_e^{})^{1/2}$ to denote the particular branch $(I_e^{})^{1/2}\equiv |U_{e1}^{}||U_{e2}^{}||U_{e3}^{}|\Delta_{12}^{}\Delta_{23}^{}\Delta_{31}^{}$, and $(\widetilde I_e^{})^{1/2}$ is defined analogously in matter. The two independent matter invariants in Eqs.~(\ref{eq:NaumovRelationIntro}) and (\ref{eq:ElectronRowInvariantIntro}) have two salient features. First, they are expressed entirely in terms of rephasing-invariant quantities directly connected to oscillation observables. Second, they take a compact multiplicative monomial form. Beyond these two relations, several exact sum rules and more general matter invariants have also been constructed in the three-flavor framework~\cite{Harrison:2002ee,Xing:2001bg,Xing:2000ik,Kimura:2002hb,Kimura:2002wd,Jarlskog:2004be,Xing:2005gk,Xing:2019owb}.
	
	Along a different line of thought, neutrino oscillations in matter can be described by a set of first-order differential equations with respect to the matter parameter $a$~\cite{Chiu:2017ckv,Xing:2018lob}. Taking $\widetilde\Delta_{ij}^{}$ and $|V_{\alpha i}^{}|^2$ as functions of $a$, these equations can be written explicitly as
	\begin{align}
	\frac{{\rm d}\widetilde\Delta_{ij}^{}}{{\rm d}a}
	=&|V_{ei}^{}|^2-|V_{ej}^{}|^2
	\;, \\
	\frac{{\rm d}|V_{\alpha i}^{}|^2}{{\rm d}a}
	=&2\sum_{j\neq i}
	\frac{\widetilde R_{e\alpha}^{ij}}
	{\widetilde\Delta_{ij}^{}}
	\;,
	\label{eq:threeRGE}
	\end{align}
	where $\widetilde R_{\alpha\beta}^{ij}\equiv{\rm Re}(
	V_{\alpha i}^{}V_{\beta j}^{}V_{\alpha j}^{*}V_{\beta i}^{*})$ represents the real quartet. In the standard three-flavor framework, for $\alpha\neq\beta$ and $\{i,j,k\}=\{1,2,3\}$, the orthogonality relation implies
	\begin{equation}
	2\widetilde R_{\alpha\beta}^{ij}
	=|V_{\alpha k}^{}|^2|V_{\beta k}^{}|^2
	-|V_{\alpha i}^{}|^2|V_{\beta i}^{}|^2
	-|V_{\alpha j}^{}|^2|V_{\beta j}^{}|^2
	\;.
	\label{eq:threeRfromModuli}
	\end{equation}
	Thus, Eq.~(\ref{eq:threeRGE}) forms a closed system of first-order differential equations for $\widetilde\Delta_{ij}^{}$ and $|V_{\alpha i}^{}|^2$. Treating the matter parameter as a renormalization scale, these differential equations closely resemble the renormalization group equations (RGEs) for running parameters. Indeed, this analogy with the RGEs governing the scale evolution of flavor-mixing parameters was further explored in Ref.~\cite{Xing:2018kto}. Approximate analytical solutions to the matter RGEs were subsequently obtained~\cite{Wang:2019yfp} and applied to studies of leptonic CP violation~\cite{Wang:2019dal,Petcov:2018zka}, while the formalism was also extended to the four-flavor neutrino oscillation framework~\cite{Zeng:2022rxm}.

	The RGE formulation of matter effects not only provides an equivalent description, but also renders the evolution of the effective oscillation parameters with the matter potential more transparent. In particular, it is convenient for uncovering matter invariants, as every such invariant corresponds to a first integral of the differential equations. This perspective was explored in Ref.~\cite{Zhou:2020iei}, which identified the continuous Lie-group symmetries and the discrete permutation symmetries of the matter RGEs and discussed their implications for constructing differential invariants. In the present work, we aim to address two questions: (i) whether the complete set of matter invariants can be derived directly from the full RGEs; and (ii) whether there exist any additional compact multiplicative relations between vacuum and matter analogous to the Naumov and electron-row relations, i.e., $\widetilde I_{\rm N}^{} = I_{\rm N}^{}$ in Eq.~(\ref{eq:NaumovRelationIntro}) and $\widetilde I_e^{} = I_e^{}$ in Eq.~(\ref{eq:ElectronRowInvariantIntro}), from which the Toshev relation follows. Specifically, in the three-flavor case, we exploit the exact permutation covariance under relabeling of the mass eigenstates to organize $\widetilde\Delta_{ij}^{}$ and $|V_{\alpha i}^{}|^2$ into $S^{}_3$-singlet combinations. We then examine their renormalization-group evolution and derive the complete invariant ring. By treating the flavor structure of the matter potential as a flavor-covariant matter spurion, we further show that the existence of multiplicative monomial invariants is controlled by its rank and by whether the residues of the $1/\widetilde\Delta_{ij}^{}$ poles in the RGEs allow a common cancellation. The three-flavor framework contains no multiplicative invariant constructed from $\widetilde\Delta_{ij}^{}$ and rephasing-invariant mixing quantities that is algebraically independent of $\widetilde I_{\rm N}^{}$ and $\widetilde I_e^{}$. Finally, we extend the analysis to the four-flavor framework, where we construct the complete polynomial invariant ring and show that, for a generic rank-two matter spurion, no nontrivial multiplicative monomial invariant of the above type exists.

	The remainder of this paper is organized as follows. In Sec.~\ref{sec:threeflavor}, we derive the complete three-flavor invariant ring and classify its multiplicative invariants. In Sec.~\ref{sec:four}, we extend both analyses to the four-flavor framework. Sec.~\ref{sec:summary} summarizes our main results.

	\section{Three-flavor neutrino oscillations}
	\label{sec:threeflavor}

	In this section, we combine the $S^{}_3$ permutation symmetry with the RGEs to derive the complete ring of differential invariants in the three-flavor case. These invariants are required to be rephasing invariant and insensitive to an overall trace shift of the Hamiltonian. We then identify the independent generators of the multiplicative monomial sector of this invariant ring.

	\subsection{\texorpdfstring{$S^{}_3$}{S3}-invariant variables}
	\label{subsec:S3invariants}

	The three-flavor effective Hamiltonian and the corresponding RGEs of oscillation parameters are invariant under a simultaneous $S^{}_3$ permutation of three neutrino mass-squared eigenvalues $\{\widetilde m_1^2,\widetilde m_2^2,\widetilde m_3^2\}$ and their corresponding eigenvectors $\{V_{\alpha1}^{},V_{\alpha2}^{},V_{\alpha3}^{}\}$~\cite{Zhou:2016luk,Xing:2018lob,Kuo:2018mnm,Kuo:2019psm}. Notice that $V_{\alpha i}^{}$ obey the row and column normalization conditions
	\begin{equation}
	\sum_i |V_{\alpha i}^{}|^2=1
	\;,
	\quad
	\sum_\alpha |V_{\alpha i}^{}|^2=1
	\;,
	\end{equation}
	and the three mass-squared differences $\widetilde\Delta_{ij}^{}$ satisfy
	\begin{equation}
	\widetilde\Delta_{12}^{}
	+\widetilde\Delta_{23}^{}
	+\widetilde\Delta_{31}^{}=0
	\;,
	\end{equation}
	indicating that only two mass-squared differences and four moduli are independent. 

	The triplets $(\widetilde m_1^2,\widetilde m_2^2,\widetilde m_3^2)^{\rm T}_{}$ and $(|V_{\alpha 1}^{}|^2,|V_{\alpha 2}^{}|^2,|V_{\alpha 3}^{}|^2)^{\rm T}_{}$ (for $\alpha=e,\mu,\tau$) each transform as the three-dimensional permutation representation of $S^{}_3$. This reducible representation decomposes as $\boldsymbol{3}=\boldsymbol{1}\oplus\boldsymbol{2}$. The trivial singlet is spanned by $\boldsymbol{1}\equiv(1,\,1,\,1)^{\rm T}$, whereas its orthogonal complement is a two-dimensional $S^{}_3$-covariant subspace carrying the irreducible doublet representation. More explicitly, we remove the identity shift from $\widetilde m_i^2$ by defining
	\begin{equation}
	\widetilde\ell_i^{}
	\equiv
	\widetilde m_i^2-\frac{1}{3}\sum_{j=1}^{3}\widetilde m_j^2
	\;,
	\label{eq:ellThree}
	\end{equation}
	which can be expressed entirely in terms of $\widetilde\Delta_{ij}^{}$ as
	\begin{equation}
	\widetilde\ell_1^{} = \frac{1}{3}\left(\widetilde\Delta_{12}^{}-\widetilde\Delta_{31}^{}\right)\;,
	\quad
	\widetilde\ell_2^{} = \frac{1}{3}\left(\widetilde\Delta_{23}^{}-\widetilde\Delta_{12}^{}\right)\;,
	\quad
	\widetilde\ell_3^{} = \frac{1}{3}\left(\widetilde\Delta_{31}^{}-\widetilde\Delta_{23}^{}\right)\;.
	\label{eq:ellThreeExplicit}
	\end{equation}
	Meanwhile, $|V_{\alpha i}^{}|^2$ can be reorganized as
	\begin{equation}
	r_i^e\equiv |V_{ei}^{}|^2-\frac{1}{3}
	\;,
	\quad
	r_i^\mu\equiv |V_{\mu i}^{}|^2-\frac{1}{3}
	\;,
	\label{eq:centeredRowsThree}
	\end{equation}
	and the third row is then fixed by column normalization
	\begin{equation}
	|V_{\tau i}^{}|^2-\frac{1}{3}=-r_i^e-r_i^\mu
	\;.
	\label{eq:centertau}
	\end{equation}
	By construction,
	\begin{equation}
	\sum_i\widetilde\ell_i^{}=\sum_i r_i^e=\sum_i r_i^\mu=0 \;.
	\label{eq:zerosum}
	\end{equation}
	Hence the centered triplets $(\widetilde\ell_1^{},\widetilde\ell_2^{},\widetilde\ell_3^{})^{\rm T}$, $(r_1^e,r_2^e,r_3^e)^{\rm T}$ and $(r_1^\mu,r_2^\mu,r_3^\mu)^{\rm T}$ all lie in the subspace orthogonal to \(\boldsymbol{1}\) and therefore transform as the irreducible doublet representation of $S^{}_3$.

	With the help of $\widetilde\ell_i^{}$, $r_i^e$ and $r_i^\mu$, we can construct the general form of an $S^{}_3$ invariant. For non-negative integers $(l,m,n)$, we define the symmetric sums
	\begin{equation}
	p_{lmn}^{}
	\equiv
	\sum_{i=1}^{3}
	\left(\widetilde\ell_i^{}\right)^l
	\left(r_i^e\right)^m
	\left(r_i^\mu\right)^n
	\;,
	\label{eq:S3Sums}
	\end{equation}
	which are invariant under simultaneous permutations of the indices $\{1,2,3\}$ labeling the three neutrino mass eigenstates. In the three-flavor case, the sums with $l+m+n\leq3$ suffice to generate all polynomial invariants under $S^{}_3$. Those with $l+m+n=1$ vanish by Eq.~(\ref{eq:zerosum}), leaving 16 nontrivial polynomial invariants. As discussed above, the Hamiltonian governing three-flavor neutrino oscillations in matter, however, admits a smaller set of independent invariants. In fact, the three triplets $(\widetilde\ell_i^{},r_i^e,r_i^\mu)$ contain only six independent degrees of freedom. For a non-degenerate mass spectrum, the values of $\widetilde\ell_i^{}$ serve as spectral labels that distinguish the three mass branches. Weighting the three triplets by $\widetilde\ell_i^{}$ and $(\widetilde\ell_i^{})^2$, respectively, we are led to the minimal set of six building blocks, from which $(\widetilde\ell_i^{},r_i^e,r_i^\mu)$ can be reconstructed up to a simultaneous permutation:
	\begin{align}
	\widetilde s_2^{}
	&\equiv\sum_i\left(\widetilde\ell_i^{}\right)^2
	\;,
	&
	\widetilde u_e^{}
	&\equiv\sum_i\widetilde\ell_i^{}r_i^e
	\;,
	&
	\widetilde w_e^{}
	&\equiv\sum_i\left(\widetilde\ell_i^{}\right)^2r_i^e
	\;,
	\nonumber\\
	\widetilde s_3^{}
	&\equiv\sum_i\left(\widetilde\ell_i^{}\right)^3
	\;,
	&
	\widetilde u_\mu^{}
	&\equiv\sum_i\widetilde\ell_i^{}r_i^\mu
	\;,
	&
	\widetilde w_\mu^{}
	&\equiv\sum_i\left(\widetilde\ell_i^{}\right)^2r_i^\mu
	\;.
	\label{eq:S3SeparatingVariables}
	\end{align}

	To show that the above six variables capture all information for a non-degenerate spectrum, we take $(\widetilde\ell_1^{},\widetilde\ell_2^{},r_1^e,r_2^e,r_1^\mu,r_2^\mu)$ as independent coordinates, with the third component of each triplet fixed by Eq.~(\ref{eq:zerosum}). The resulting Jacobian is
	\begin{equation}
	\det
	\frac{\partial
	\left(
	\widetilde s_2^{},\widetilde s_3^{},
	\widetilde u_e^{},\widetilde u_\mu^{},
	\widetilde w_e^{},\widetilde w_\mu^{}
	\right)}
	{\partial
	\left(
	\widetilde\ell_1^{},\widetilde\ell_2^{},
	r_1^e,r_2^e,
	r_1^\mu,r_2^\mu
	\right)}
	=6
	\left(\widetilde\ell_1^{}-\widetilde\ell_2^{}\right)^3
	\left(\widetilde\ell_2^{}-\widetilde\ell_3^{}\right)^3
	\left(\widetilde\ell_1^{}-\widetilde\ell_3^{}\right)^3
	\;,
	\label{eq:S3SeparatingJacobian}
	\end{equation}
	which is nonzero whenever the matter spectrum is non-degenerate, so the six variables are mutually independent. Their completeness can be seen by reconstructing the original triplets. First, $\widetilde s_2^{}$ and $\widetilde s_3^{}$ determine $\widetilde\ell_i^{}$ as the roots of a cubic equation
	\begin{equation}
	x^3-\frac{\widetilde s_2^{}}{2}x
	-\frac{\widetilde s_3^{}}{3}=0
	\;.
	\label{eq:S3CubicSpectrum}
	\end{equation}
	Second, each centered flavor triplet can be found by inverting the Vandermonde matrix, namely,
	\begin{equation}
	\begin{pmatrix}
		1 & 1 & 1\\
		\widetilde\ell_1^{} & \widetilde\ell_2^{} & \widetilde\ell_3^{}\\
		\left(\widetilde\ell_1^{}\right)^2 & \left(\widetilde\ell_2^{}\right)^2 & \left(\widetilde\ell_3^{}\right)^2
	\end{pmatrix}
	\begin{pmatrix}
		r_1^\alpha\\
		r_2^\alpha\\
		r_3^\alpha
	\end{pmatrix}
	=
	\begin{pmatrix}
		0\\
		\widetilde u_\alpha^{}\\
		\widetilde w_\alpha^{}
	\end{pmatrix}
	\;,
	\quad
	\alpha=e,\mu
	\;.
	\label{eq:S3VandermondeReconstruction}
	\end{equation}
	The six combinations in Eq.~(\ref{eq:S3SeparatingVariables}) therefore generate the $S^{}_3$-invariant rational function field when the three neutrino mass eigenvalues are non-degenerate, and retain all information contained in $\widetilde\Delta_{ij}^{}$ and $|V_{\alpha i}^{}|^2$ up to a simultaneous $S^{}_3$ relabeling of the neutrino mass eigenstates.

	\subsection{Exact first integrals from the RGEs}
	\label{subsec:S3InducedRGE}

	From Eq.~(\ref{eq:threeRGE}), the variables $\widetilde\ell_i^{}$ obey
	\begin{equation}
	\frac{{\rm d}\widetilde\ell_i^{}}{{\rm d}a}=r_i^e
	\;.
	\label{eq:ellRGEThree}
	\end{equation}
	Then, for the variables defined in Eq.~(\ref{eq:S3SeparatingVariables}), we obtain the following RGEs
	\begin{align}
	\frac{{\rm d}\widetilde s_2^{}}{{\rm d}a}
	&=2\widetilde u_e^{}
	\;,
	&
	\frac{{\rm d}\widetilde u_e^{}}{{\rm d}a}
	&=\frac{2}{3}
	\;,
	&
	\frac{{\rm d}\widetilde w_e^{}}{{\rm d}a}
	&=\frac{2}{3}\widetilde u_e^{}
	\;,
	\nonumber\\
	\frac{{\rm d}\widetilde s_3^{}}{{\rm d}a}
	&=3\widetilde w_e^{}
	\;,
	&
	\frac{{\rm d}\widetilde u_\mu^{}}{{\rm d}a}
	&=-\frac{1}{3}
	\;,
	&
	\frac{{\rm d}\widetilde w_\mu^{}}{{\rm d}a}
	&=-\frac{2}{3}
	\left(\widetilde u_e^{}+\widetilde u_\mu^{}\right)
	\;.
	\label{eq:S3InducedRGE}
	\end{align}
	It is worth noting that all rational poles $\propto 1/\widetilde\Delta_{ij}^{}$ in the RGEs for $|V_{\alpha i}^{}|^2$ cancel out in the above differential equations, leaving a purely polynomial system. The derivation of these equations is straightforward. For example,
	\begin{align}
	\frac{{\rm d}\widetilde u_e^{}}{{\rm d}a}
	&=
	\frac{{\rm d}}{{\rm d}a}
	\sum_i\widetilde\ell_i^{}r_i^e
	\nonumber\\
	&=
	\sum_i\left(|V_{ei}^{}|^2-\frac{1}{3}\right)|V_{ei}^{}|^2
	+2\sum_{i<j}|V_{ei}^{}|^2|V_{ej}^{}|^2
	\nonumber\\
	&=\left(\sum_i |V_{ei}^{}|^2\right)^2-\frac{1}{3}
	=\frac{2}{3}
	\;,
	\label{eq:ueRGEDerivation}
	\end{align}
	where the pole terms in the moduli RGEs are canceled pairwise, and row normalization has been used to derive the final result. For $\widetilde u_\mu^{}$, orthogonality of the $e$ and $\mu$ rows instead gives ${\rm d}\widetilde u_\mu^{}/{\rm d}a=-1/3$. The remaining four equations follow from the same pairwise cancellation together with row and column unitarity.

	The six variables depend on a single matter parameter $a$, hence five first integrals are expected, which can be constructed systematically order by order in $\widetilde\ell_i$. One convenient polynomial set is
	\begin{align}
	\widetilde{\cal I}_1^{}
	&\equiv
	\widetilde u_e^{}+2\widetilde u_\mu^{}
	\;,
	\label{eq:S3I1}\\
	\widetilde{\cal I}_2^{}
	&\equiv
	\frac{\widetilde s_2^{}}{6}
	+\widetilde w_e^{}+\widetilde w_\mu^{}
	+\widetilde u_e^{}\widetilde u_\mu^{}
	\;,
	\label{eq:S3I2}\\
	\widetilde{\cal I}_3^{}
	&\equiv
	\frac{\widetilde s_2^{}}{6}
	-\widetilde w_\mu^{}
	-\widetilde u_e^2
	-\widetilde u_e^{}\widetilde u_\mu^{}
	\;,
	\label{eq:S3I3}\\
	\widetilde{\cal I}_4^{}
	&\equiv
	\frac{\widetilde s_2^{}}{6}
	-\widetilde w_e^{}
	-\widetilde u_e^{}\widetilde u_\mu^{}
	-\widetilde u_\mu^2
	\;,
	\label{eq:S3I4}\\
	\widetilde{\cal I}_5^{}
	&\equiv
	\frac{\widetilde s_3^{}}{6}
	-\widetilde u_e^{}\widetilde u_\mu^{}
	\left(\widetilde u_e^{}+\widetilde u_\mu^{}\right)
	-\widetilde u_e^{}\widetilde w_e^{}
	-\widetilde u_\mu^{}\widetilde w_\mu^{}
	-\frac{1}{2}\widetilde u_e^{}\widetilde w_\mu^{}
	-\frac{1}{2}\widetilde u_\mu^{}\widetilde w_e^{}
	\;.
	\label{eq:S3I5}
	\end{align}
	Direct substitution of Eq.~(\ref{eq:S3InducedRGE}) gives
	\begin{equation}
	\frac{{\rm d}\widetilde{\cal I}_A^{}}{{\rm d}a}=0
	\;,
	\quad
	A=1,\ldots,5
	\;.
	\label{eq:S3IConstant}
	\end{equation}
	Thus each $\widetilde{\cal I}_A^{}$ is an exact matter invariant. The independence and completeness of these five invariants can be established by checking the following Jacobian
	\begin{equation}
	\det
	\frac{\partial
	\left(
	\widetilde u_e^{},
	\widetilde{\cal I}_1^{},\ldots,
	\widetilde{\cal I}_5^{}
	\right)}
	{\partial
	\left(
	\widetilde s_2^{},\widetilde s_3^{},
	\widetilde u_e^{},\widetilde u_\mu^{},
	\widetilde w_e^{},\widetilde w_\mu^{}
	\right)}
	=-\frac{1}{6}
	\;,
	\label{eq:S3TriangularJacobian}
	\end{equation}
	which shows that the transformation is locally invertible.
	Only $\widetilde u_e^{}$ evolves with $a$, whereas the five transverse coordinates $\widetilde{\cal I}_A^{}$ remain fixed. Therefore, Eqs.~(\ref{eq:S3I1})--(\ref{eq:S3I5}) constitute a complete set of five independent first integrals. Once their vacuum values are specified, these relations impose five exact constraints on $\widetilde\Delta_{ij}^{}$ and $|V_{\alpha i}^{}|^2$ at any matter potential.

	\subsection{Physical identification of the invariants}
	\label{subsec:threePhysicalRing}
	The five invariants derived in the previous subsection can be rewritten in terms of $\widetilde\Delta_{ij}^{}$ and $|V_{\alpha i}^{}|^2$. Substituting the definitions in Eqs.~(\ref{eq:centeredRowsThree}) and~(\ref{eq:S3SeparatingVariables}) into Eqs.~(\ref{eq:S3I1})--(\ref{eq:S3I5}), and then using Eq.~(\ref{eq:ellThreeExplicit}) together with column unitarity, we obtain
	\begin{align}
	\widetilde{\cal I}_1^{}
	&=
	\frac{1}{3}
	\sum_{i=1}^{3}
	\left(
	\widetilde\Delta_{i,i+1}^{}
	-\widetilde\Delta_{i-1,i}^{}
	\right)
	\left(
	|V_{\mu i}^{}|^2
	-|V_{\tau i}^{}|^2
	\right)
	\;,
	\label{eq:S3I1MassModuli}\\
	\widetilde{\cal I}_2^{}
	&=
	\frac{1}{9}
	\Biggl\{
	\sum_{i=1}^{3}
	\left(
	\widetilde\Delta_{i,i+1}^{}
	-\widetilde\Delta_{i-1,i}^{}
	\right)^2
	\left(
	\frac{1}{2}-|V_{\tau i}^{}|^2
	\right)
	\nonumber\\
	&\hspace{1cm}
	+\left[
	\sum_{i=1}^{3}
	\left(
	\widetilde\Delta_{i,i+1}^{}
	-\widetilde\Delta_{i-1,i}^{}
	\right)
	|V_{e i}^{}|^2
	\right]
	\left[
	\sum_{j=1}^{3}
	\left(
	\widetilde\Delta_{j,j+1}^{}
	-\widetilde\Delta_{j-1,j}^{}
	\right)
	|V_{\mu j}^{}|^2
	\right]
	\Biggr\}
	\;,
	\label{eq:S3I2MassModuli}\\
	\widetilde{\cal I}_3^{}
	&=
	\frac{1}{9}
	\Biggl\{
	\sum_{i=1}^{3}
	\left(
	\widetilde\Delta_{i,i+1}^{}
	-\widetilde\Delta_{i-1,i}^{}
	\right)^2
	\left(
	\frac{1}{2}-|V_{\mu i}^{}|^2
	\right)
	\nonumber\\
	&\hspace{1cm}
	+\left[
	\sum_{i=1}^{3}
	\left(
	\widetilde\Delta_{i,i+1}^{}
	-\widetilde\Delta_{i-1,i}^{}
	\right)
	|V_{e i}^{}|^2
	\right]
	\left[
	\sum_{j=1}^{3}
	\left(
	\widetilde\Delta_{j,j+1}^{}
	-\widetilde\Delta_{j-1,j}^{}
	\right)
	|V_{\tau j}^{}|^2
	\right]
	\Biggr\}
	\;,
	\label{eq:S3I3MassModuli}\\
	\widetilde{\cal I}_4^{}
	&=
	\frac{1}{9}
	\Biggl\{
	\sum_{i=1}^{3}
	\left(
	\widetilde\Delta_{i,i+1}^{}
	-\widetilde\Delta_{i-1,i}^{}
	\right)^2
	\left(
	\frac{1}{2}-|V_{e i}^{}|^2
	\right)
	\nonumber\\
	&\hspace{1cm}
	+\left[
	\sum_{i=1}^{3}
	\left(
	\widetilde\Delta_{i,i+1}^{}
	-\widetilde\Delta_{i-1,i}^{}
	\right)
	|V_{\mu i}^{}|^2
	\right]
	\left[
	\sum_{j=1}^{3}
	\left(
	\widetilde\Delta_{j,j+1}^{}
	-\widetilde\Delta_{j-1,j}^{}
	\right)
	|V_{\tau j}^{}|^2
	\right]
	\Biggr\}
	\;,
	\label{eq:S3I4MassModuli}\\
	\widetilde{\cal I}_5^{}
	&=
	\frac{1}{54}
	\Biggl\{
	\frac{1}{3}
	\sum_{i=1}^{3}
	\left(
	\widetilde\Delta_{i,i+1}^{}
	-\widetilde\Delta_{i-1,i}^{}
	\right)^3
	+2
	\prod_{\alpha=e,\mu,\tau}
	\left[
	\sum_{i=1}^{3}
	\left(
	\widetilde\Delta_{i,i+1}^{}
	-\widetilde\Delta_{i-1,i}^{}
	\right)
	|V_{\alpha i}^{}|^2
	\right]
	\nonumber\\
	&\hspace{1cm}
	-\sum_{\alpha=e,\mu,\tau}
	\left[
	\sum_{i=1}^{3}
	\left(
	\widetilde\Delta_{i,i+1}^{}
	-\widetilde\Delta_{i-1,i}^{}
	\right)
	|V_{\alpha i}^{}|^2
	\right]
	\left[
	\sum_{j=1}^{3}
	\left(
	\widetilde\Delta_{j,j+1}^{}
	-\widetilde\Delta_{j-1,j}^{}
	\right)^2
	|V_{\alpha j}^{}|^2
	\right]
	\Biggr\}
	\;,
	\label{eq:S3I5MassModuli}
	\end{align}
	where the mass indices in these expressions should be understood cyclically modulo three. 

	The complete set of independent matter invariants for three-flavor neutrino oscillations has been characterized previously through complementary approaches. These include a direct construction based on the matter-independent $H_{\alpha\beta}^{}$ for $(\alpha,\beta)\neq(e,e)$ combined with flavor-rephasing invariance~\cite{Harrison:2002ee}, the matter-invariant commutator of the lepton mass matrices~\cite{Xing:2001bg,Xing:2000ik,Harrison:1999df}, Vandermonde identities for the effective mass eigenvalues~\cite{Kimura:2002hb,Kimura:2002wd,Jarlskog:2004be,Xing:2005gk,Xing:2019owb}, and, more recently, an adjugate-based formulation of the Hamiltonian~\cite{Abdullahi:2022fkh}. It is therefore useful to compare the five invariants obtained above with the previously established results and to make explicit how the corresponding invariant bases are related.

	In Ref.~\cite{Harrison:2002ee}, after removing the unobservable trace and flavor-rephasing phases, the diagonal invariant is given by
	\begin{align}
	\widetilde D_{\mu\tau}^{}
	\equiv
	M_{\mu\mu}^{}(a)-M_{\tau\tau}^{}(a)
	=
	\widetilde\Delta_{21}^{}
	\left(
	|V_{\mu2}^{}|^2
	-|V_{\tau2}^{}|^2
	\right)
	+
	\widetilde\Delta_{31}^{}
	\left(
	|V_{\mu3}^{}|^2
	-|V_{\tau3}^{}|^2
	\right)	
	\;,
	\label{eq:S3DiagonalIdentification}
	\end{align}
	where $M(a)\equiv2EH(a)$. This invariant is equivalent to $\widetilde{\cal I}_1^{}$ in Eq.~(\ref{eq:S3I1MassModuli}). For the off-diagonal sector, define the matter covariants~\cite{Xing:2001bg}
	\begin{equation}
	\widetilde Z_{\alpha\beta}^{}
	\equiv
	\sum_{i=1}^{3}\widetilde m_i^2
	V_{\alpha i}^{}V_{\beta i}^{*}
	\;,
	\quad
	\alpha\neq\beta
	\;,
	\label{eq:Zthree}
	\end{equation}
	and let $Z_{\alpha\beta}^{}\equiv\sum_i m_i^2U_{\alpha i}^{}U_{\beta i}^{*}$ denote their vacuum counterparts. Since the matter potential is diagonal in the flavor basis, $\widetilde Z_{\alpha\beta}^{}$ is independent of $a$, leading to the following sum rules~\cite{Xing:2001bg,Xing:2003ez,Xing:2005gk,Xing:2019owb}
	\begin{equation}
	\sum_{i=1}^{3}\widetilde m_i^2
	V_{\alpha i}^{}V_{\beta i}^{*}
	=
	\sum_{i=1}^{3} m_i^2
	U_{\alpha i}^{}U_{\beta i}^{*}
	\;.
	\label{eq:nondiagsumrules}
	\end{equation}
	Taking the modulus squared of Eq.~(\ref{eq:nondiagsumrules}) removes the flavor-rephasing phase and gives the exact matter invariants depending only on $\widetilde\Delta_{ij}^{}$ and $|V_{\alpha i}^{}|^2$~\cite{Harrison:2002ee}, namely,
	\begin{equation}
	\widetilde B_{\alpha\beta}^{}\equiv|\widetilde Z_{\alpha\beta}^{}|^2
	= -\sum_{i<j}
	\widetilde\Delta_{ij}^{2}
	\widetilde R_{\alpha\beta}^{ij} \;,
	\label{eq:Binvariant}
	\end{equation}
	where the real quartet $\widetilde R_{\alpha\beta}^{ij}\equiv{\rm Re}(V_{\alpha i}^{}V_{\beta j}^{}V_{\alpha j}^{*}V_{\beta i}^{*})$ can be expressed entirely in terms of $|V_{\alpha i}^{}|^2$ in the three-flavor case, as shown in Eq.~(\ref{eq:threeRfromModuli}). To see how $\widetilde m_i^2$ in Eq.~(\ref{eq:nondiagsumrules}) combine into $\widetilde\Delta_{ij}^{}$ in Eq.~(\ref{eq:Binvariant}), notice that row orthogonality gives $\sum_{j\neq i}\widetilde R_{\alpha\beta}^{ij}=-|V_{\alpha i}^{}|^2|V_{\beta i}^{}|^2$ for $\alpha\neq\beta$. Using this identity to eliminate the diagonal terms yields
	\begin{align}
	\widetilde B_{\alpha\beta}^{}
	&=
	\sum_i
	\widetilde m_i^4
	|V_{\alpha i}^{}|^2
	|V_{\beta i}^{}|^2
	+2\sum_{i<j}
	\widetilde m_i^2\widetilde m_j^2
	\widetilde R_{\alpha\beta}^{ij}
	\nonumber\\
	&=-\sum_{i<j}
	\left(\widetilde m_i^2-\widetilde m_j^2\right)^2
	\widetilde R_{\alpha\beta}^{ij}
	=-\sum_{i<j}
	\widetilde\Delta_{ij}^{2}
	\widetilde R_{\alpha\beta}^{ij}
	\;.
	\label{eq:Bthree}
	\end{align}
	Then one can check that three quadratic RGE invariants $\{\widetilde{\cal I}_2^{},\widetilde{\cal I}_3^{},\widetilde{\cal I}_4^{}\}$ precisely correspond to the three sum rules as
	\begin{equation}
	\widetilde{\cal I}_2^{}
	=\widetilde B_{e\mu}^{}
	\;,
	\quad
	\widetilde{\cal I}_3^{}
	=\widetilde B_{e\tau}^{}
	\;,
	\quad
	\widetilde{\cal I}_4^{}
	=\widetilde B_{\mu\tau}^{}
	\;.
	\label{eq:S3PhysicalIdentification}
	\end{equation}

	The remaining rephasing-invariant information in the off-diagonal sector is contained in the closed product
	\begin{equation}
	\widetilde C^{}
	\equiv
	\widetilde Z_{e\mu}^{}
	\widetilde Z_{\mu\tau}^{}
	\widetilde Z_{\tau e}^{}
	\;.
	\label{eq:Cthree}
	\end{equation}
	We use $\widetilde C_{\rm R}^{}$ and $\widetilde C_{\rm I}^{}$ to denote the real and imaginary parts of $\widetilde C^{}$, respectively. $\widetilde C_{\rm I}^{}$ gives rise to the Naumov relation
	\begin{equation}
	\widetilde C_{\rm I}^{}
	=
	\widetilde J\,
	\widetilde\Delta_{12}^{}
	\widetilde\Delta_{23}^{}
	\widetilde\Delta_{31}^{}
	=
	J\,\Delta_{12}^{}\Delta_{23}^{}\Delta_{31}^{}
	=C_{\rm I}^{}
	\;.
	\label{eq:CthreeNaumov}
	\end{equation}
	The last invariant derived from the RGEs instead selects the real part, $\widetilde{\cal I}_5^{}
	=\widetilde C_{\rm R}^{}$, which is the main distinction between the present construction and the particular complete set chosen in Ref.~\cite{Harrison:2002ee}. The latter uses $\widetilde C_{\rm I}^{}$ as the remaining off-diagonal invariant, whereas the RGEs for $|V_{\alpha i}^{}|^2$ and $\widetilde\Delta_{ij}^{}$ select $\widetilde C_{\rm R}^{}$, which is manifestly CP even. The CP-odd invariant $\widetilde C_{\rm I}^{}$ satisfies the algebraic relation
	\begin{align}
	\widetilde C_{\rm I}^2
	=
	\widetilde B_{e\mu}^{}
	\widetilde B_{e\tau}^{}
	\widetilde B_{\mu\tau}^{}
	-\widetilde C_{\rm R}^2
	=
	\widetilde{\cal I}_2^{}
	\widetilde{\cal I}_3^{}
	\widetilde{\cal I}_4^{}
	-\widetilde{\cal I}_5^2
	\;,
	\label{eq:CIthreeSquare}
	\end{align}
	which only fixes the modulus of $\widetilde C_{\rm I}^{}$. Recovering the exact Naumov relation in Eq.~(\ref{eq:CthreeNaumov}) requires the sign of $\widetilde J$.

	Now we can write down the complete polynomial ring in the three-flavor case as
	\begin{equation}
	\widetilde\Rcal_{3\nu}^{}
	=
	\mathbb R[
	\widetilde D_{\mu\tau}^{},
	\widetilde B_{e\mu}^{},
	\widetilde B_{e\tau}^{},
	\widetilde B_{\mu\tau}^{},
	\widetilde C_{\rm R}^{},
	\widetilde C_{\rm I}^{}]
	\big/
	\left\langle
	\widetilde C_{\rm R}^{2}
	+\widetilde C_{\rm I}^{2}
	-\widetilde B_{e\mu}^{}
	\widetilde B_{e\tau}^{}
	\widetilde B_{\mu\tau}^{}
	\right\rangle
	\;.
	\label{eq:threeSignedRing}
	\end{equation}
	Eq.~(\ref{eq:threeSignedRing}) provides a polynomial description of the rephasing-invariant quantities and retains the sign of the CP-odd invariant $\widetilde C_{\rm I}^{}$. The invariant ring has six real generators subject to one algebraic relation, leaving five algebraically independent degrees of freedom. The RGE derivation above and the direct Hamiltonian construction therefore give the same counting and a mutually consistent identification of the complete set of three-flavor matter invariants.

	Now that the complete invariant ring has been constructed, it is worth checking whether the Toshev relation can also be derived from the ring generators. In the following, we show that the Toshev invariant satisfies $\widetilde I_{\rm T}^{}=\widetilde I_{\rm N}^{}/(\widetilde I_e^{})^{1/2}$ with $\widetilde I_{\rm N}^{}=\widetilde C_{\rm I}^{}$, and $\widetilde I_e^{}$ is given in terms of the ring generators. We introduce the following matrix
	\begin{equation}
	\mathcal K_e^{}
	\equiv
	\begin{pmatrix}
	\langle e|\\
	\langle e|M\\
	\langle e|M^2
	\end{pmatrix}
	\;,
	\label{eq:electron}
	\end{equation}
	where $\langle e|=(1,0,0)$ in the flavor basis. Since $M=V\diag(\widetilde m_1^2,\widetilde m_2^2,\widetilde m_3^2)V^\dagger$, we have
	\begin{equation}
	\mathcal K_e^{}V
	=
	\begin{pmatrix}
	V_{e1}^{} & V_{e2}^{} & V_{e3}^{}\\
	\widetilde m_1^2V_{e1}^{} & \widetilde m_2^2V_{e2}^{} & \widetilde m_3^2V_{e3}^{}\\
	\widetilde m_1^4V_{e1}^{} & \widetilde m_2^4V_{e2}^{} & \widetilde m_3^4V_{e3}^{}
	\end{pmatrix}
	\;.
	\end{equation}
	Given that $|\det V|=1$, the Vandermonde determinant then gives
	\begin{equation}
	\left|\det\mathcal K_e^{}\right|^2
	=
	|V_{e1}^{}V_{e2}^{}V_{e3}^{}|^2
	\prod_{i<j}
	\left(\widetilde m_i^2-\widetilde m_j^2\right)^2
	=
	|V_{e1}^{}|^2
	|V_{e2}^{}|^2
	|V_{e3}^{}|^2
	\widetilde\Delta_{12}^2
	\widetilde\Delta_{23}^2
	\widetilde\Delta_{31}^2
	\equiv
	\widetilde I_e^{}
	\;.
	\label{eq:IeMass}
	\end{equation}
	On the other hand, the same determinant can be calculated in the flavor basis. Using $M_{\alpha\beta}^{}=\widetilde Z_{\alpha\beta}^{}$ for $\alpha\neq\beta$ and $M_{\mu\mu}^{}-M_{\tau\tau}^{}=\widetilde D_{\mu\tau}^{}$, we find
	\begin{align}
	\det\mathcal K_e^{}
	=
	\widetilde Z_{e\mu}^{}(M^2)_{e\tau}^{}
	-\widetilde Z_{e\tau}^{}(M^2)_{e\mu}^{}
	=
	\widetilde Z_{e\mu}^{2}\widetilde Z_{\mu\tau}^{}
	-\widetilde Z_{e\tau}^{2}\widetilde Z_{\mu\tau}^{*}
	-\widetilde D_{\mu\tau}^{}
	\widetilde Z_{e\mu}^{}\widetilde Z_{e\tau}^{}
	\;.
	\label{eq:IeFlavor}
	\end{align}
	It follows that
	\begin{equation}
	\widetilde I_e^{}
	=
	\left|
	\widetilde Z_{e\mu}^{2}\widetilde Z_{\mu\tau}^{}
	-\widetilde Z_{e\tau}^{2}\widetilde Z_{\mu\tau}^{*}
	-\widetilde D_{\mu\tau}^{}
	\widetilde Z_{e\mu}^{}\widetilde Z_{e\tau}^{}
	\right|^2
	\;.
	\label{eq:IeCompact}
	\end{equation}
	Expanding the modulus squared and using the algebraic relation in Eq.~(\ref{eq:threeSignedRing}), we obtain
	\begin{align}
	\widetilde I_e^{}
	=
	\widetilde D_{\mu\tau}^{2}
	\widetilde B_{e\mu}^{}\widetilde B_{e\tau}^{}
	+\widetilde B_{\mu\tau}^{}
	\left(
	\widetilde B_{e\mu}^{}-\widetilde B_{e\tau}^{}
	\right)^2
	+2\widetilde D_{\mu\tau}^{}
	\left(
	\widetilde B_{e\tau}^{}-\widetilde B_{e\mu}^{}
	\right)
	\widetilde C_{\rm R}^{}
	+4\widetilde C_{\rm I}^{2}
	\;,
	\label{eq:IeRing}
	\end{align}
	which, together with $\widetilde I_{\rm T}^{}=\widetilde I_{\rm N}^{}/(\widetilde I_e^{})^{1/2}$, establishes the explicit relation between $\widetilde I_{\rm T}^{}$ and the generators of the invariant ring.

	\subsection{Multiplicative invariants}
	We have derived from the RGEs a complete set of independent matter invariants, all of which are polynomial functions of $\widetilde\Delta_{ij}^{}$ and $|V_{\alpha i}^{}|^2$. We have also shown that compact multiplicative relations, exemplified by the Naumov and Toshev relations, follow from the invariant ring. In this subsection, we address the following question: does the three-flavor framework admit any additional independent multiplicative invariants?

	We first organize $\widetilde\Delta_{ij}^{}$ and $|V_{\alpha i}^{}|^2$ into $S^{}_3$ singlets under simultaneous permutations of the matter-eigenstate labels
	\begin{align}
	\widetilde{\cal V}_3^{}
	\equiv
	\widetilde\Delta_{12}^{}
	\widetilde\Delta_{23}^{}
	\widetilde\Delta_{31}^{}
	\;, \quad
	\widetilde\Pi_\alpha^{}
	\equiv
	\prod_{i=1}^{3}
	|V_{\alpha i}^{}|^2
	\;,
	\label{eq:VandermondeFlavorProducts}
	\end{align}
	with $\alpha=e,\mu,\tau$. For any group element $\pi\in S^{}_3$, these two quantities and the Jarlskog invariant transform as
	\begin{equation}
	\widetilde{\cal V}_3^{}
	\xrightarrow{\pi}
	{\rm sgn}(\pi)\widetilde{\cal V}_3^{}
	\;,
	\quad
	\widetilde J
	\xrightarrow{\pi}
	{\rm sgn}(\pi)\widetilde J
	\;,
	\quad
	\widetilde\Pi_\alpha^{}
	\xrightarrow{\pi}
	\widetilde\Pi_\alpha^{}
	\;.
	\label{eq:S3MultiplicativeCharacters}
	\end{equation}
	Thus $\widetilde{\cal V}_3^{}$ and $\widetilde J$ have the charge $-1$ under an odd permutation of $S^{}_3$, whereas all three row products have the charge $+1$. Interestingly, $\widetilde{\cal V}_3^{}$, $\widetilde\Pi_e^{}$, and $\widetilde J$ obey the RGEs
	\begin{equation}
	\frac{{\rm d}\ln\widetilde{\cal V}_3^{}}{{\rm d}a}
	=\widetilde\Omega^{}
	\;,
	\quad
	\frac{{\rm d}\ln\widetilde\Pi_e^{}}{{\rm d}a}
	=-2\widetilde\Omega^{}
	\;,
	\quad
	\frac{{\rm d}\ln\widetilde J}{{\rm d}a}
	=-\widetilde\Omega^{}
	\;,
	\label{eq:CommonCofactorThree}
	\end{equation}
	where the common cofactor $\widetilde\Omega^{}$ is given by
	\begin{equation}
	\widetilde\Omega^{}
	\equiv
	\frac{
	|V_{e1}^{}|^2-|V_{e2}^{}|^2}
	{\widetilde\Delta_{12}^{}}
	+
	\frac{
	|V_{e2}^{}|^2-|V_{e3}^{}|^2}
	{\widetilde\Delta_{23}^{}}
	+
	\frac{
	|V_{e3}^{}|^2-|V_{e1}^{}|^2}
	{\widetilde\Delta_{31}^{}}
	\;.
	\label{eq:OmegaThree}
	\end{equation}
	The common cofactor originates from the structure of the rank-one matter spurion $P_e^{}=|e\rangle\langle e|$. In the mass basis, $P_e^{}$ factorizes as $(V^\dagger P_e^{}V)_{ij}^{}=V_{ei}^{*}V_{ej}^{}$, indicating that the matter background ``sees'' each mass eigenstate only through its electron-flavor projection. Consequently, the RGEs of $\widetilde\Delta_{ij}^{}$ and $|V_{ei}^{}|^2$ are controlled by the same electron-flavor weights $|V_{ei}^{}|^2$, producing the common cofactor $\widetilde\Omega^{}$ in the first two relations of Eq.~(\ref{eq:CommonCofactorThree}). The Jarlskog invariant $\widetilde J$ is an exception, as it does not belong to the closed subsystem formed by $\widetilde\Delta_{ij}^{}$ and $|V_{ei}^{}|^2$. It nevertheless shares the same cofactor because every CP-odd rephasing invariant in the three-flavor case is proportional to the unique $\widetilde J$, so its RGE cannot mix with an independent CP-odd quantity and therefore must close multiplicatively on $\widetilde J$. Direct calculation then fixes its cofactor to $-\widetilde\Omega^{}$, yielding the third relation in Eq.~(\ref{eq:CommonCofactorThree}).

	As Eq.~(\ref{eq:OmegaThree}) shows, $\widetilde\Omega^{}$ consists entirely of simple $1/\widetilde\Delta_{ij}^{}$ poles and is therefore generically divergent when two effective mass eigenvalues in matter become degenerate. Because all three quantities share this single cofactor, these poles can be canceled in a multiplicative combination whose total coefficient of $\widetilde\Omega^{}$ vanishes. 
	
	By contrast, the $\mu$- and $\tau$-row products satisfy
	\begin{align}
	\frac{{\rm d}\ln\widetilde\Pi_\alpha^{}}{{\rm d}a}
	&=
	\widetilde\Omega_\alpha^{}
	\;,
	\label{eq:MuTauRGE}
	\end{align}
	with $\alpha=\mu,\tau$ and
	\begin{align}
	\widetilde\Omega_\alpha^{}
	&\equiv
	2\sum_{i<j}
	\frac{\widetilde R_{e\alpha}^{ij}}
	{\widetilde\Delta_{ij}^{}}
	\left(
	\frac{1}{|V_{\alpha i}^{}|^2}
	-\frac{1}{|V_{\alpha j}^{}|^2}
	\right)
	\;.
	\label{eq:MuTauCofactors}
	\end{align}
	The cofactors $\widetilde\Omega_\mu^{}$ and $\widetilde\Omega_\tau^{}$ contain pole residues independent of $\widetilde\Omega^{}$, which also has a direct spurion interpretation. The electron row is aligned with the one-dimensional direction selected by $P_e^{}$, so $V_{ej}^{}(V^\dagger P_e^{}V)_{ji}^{}=|V_{ej}^{}|^2V_{ei}^{}$ factorizes. For $\alpha=\mu,\tau$, the analogous contraction $V_{\alpha j}^{}(V^\dagger P_e^{}V)_{ji}^{}=V_{\alpha j}^{}V_{ej}^{*}V_{ei}^{}$ is not proportional to $V_{\alpha i}^{}$, leaving the row-dependent quartets $\widetilde R_{e\alpha}^{ij}$ in $\widetilde\Omega_\alpha^{}$. Although $P_e^{}$ treats $\mu$ and $\tau$ symmetrically, it selects no direction within the orthogonal $\mu$--$\tau$ plane. For a generic $V$, covariance under rotations in this plane does not require $\widetilde\Omega_\mu^{}=\widetilde\Omega_\tau^{}$.\footnote{Such an equality can arise only when $V$ obeys an additional constraint such as $\mu$--$\tau$ reflection symmetry~\cite{Harrison:2002et,Xing:2010ez}.} When $V_{\alpha i}^{}\to0$, one has $\widetilde R_{e\alpha}^{ij}={\cal O}(|V_{\alpha i}^{}|)$, so the corresponding term in $\widetilde\Omega_\alpha^{}$ diverges as $1/|V_{\alpha i}^{}|$, different from the $1/\widetilde\Delta_{ij}^{}$ divergence.

	Within the system of $\widetilde{\cal V}_3^{}$, $\widetilde\Pi_\alpha^{}$ and $\widetilde J$, the most general rephasing- and $S^{}_3$-invariant monomials take the form
	\begin{equation}
	\widetilde{\cal M}
	=
	\widetilde J^p
	\widetilde\Pi_e^q
	\widetilde\Pi_\mu^r
	\widetilde\Pi_\tau^s
	\widetilde{\cal V}_3^t
	\;,
	\quad
	p,q,r,s,t\in\mathbb N
	\;.
	\label{eq:ThreeFullMonomial}
	\end{equation}
	With the help of Eqs.~(\ref{eq:CommonCofactorThree}) and~(\ref{eq:MuTauRGE}), we obtain the logarithmic derivative of $\widetilde{\cal M}$
	\begin{equation}
	\frac{{\rm d}\ln\widetilde{\cal M}}{{\rm d}a}
	=
	\left(t-p-2q\right)\widetilde\Omega^{}
	+r\widetilde\Omega_\mu^{}
	+s\widetilde\Omega_\tau^{}
	\;.
	\label{eq:ThreeFullMonomialRGE}
	\end{equation}
	An exact matter invariant must cancel every independent pole residue, which can be used to determine the constraints on the exponents. The distinct singularities in Eq.~(\ref{eq:MuTauCofactors}) first require $r=s=0$, after which cancellation of the common electron-spurion cofactor gives
	\begin{equation}
	r=s=0
	\;,
	\quad
	t=p+2q
	\;.
	\label{eq:ThreeMonomialConstraints}
	\end{equation}
	Consequently, we arrive at
	\begin{equation}
	\widetilde{\cal M}
	=
	\left(
	\widetilde J\widetilde{\cal V}_3^{}
	\right)^p
	\left[
	\widetilde\Pi_e^{}
	\widetilde{\cal V}_3^2
	\right]^q
	=
	\widetilde I_{\rm N}^p
	\widetilde I_e^q
	\;.
	\label{eq:GeneralMonomialInvariant}
	\end{equation}
	Therefore, in the three-flavor case we only have two independent monomial invariants, which are precisely the Naumov invariant $\widetilde I_{\rm N}^{}$ and the electron-row invariant $\widetilde I_e^{}$. As already shown in the previous subsection, the Toshev invariant $\widetilde I_{\rm T}^{}=\widetilde I_{\rm N}^{}/(\widetilde I_e^{})^{1/2}$ belongs to their square-root extension and supplies no additional independent generator.

	\section{Four-flavor neutrino oscillations}
	\label{sec:four}

	We now extend the construction to the four-flavor framework with three active neutrinos and one sterile neutrino $\nu_s^{}$. The Hamiltonian in the flavor basis becomes
	\begin{align}
	H(a)
	=
	\frac{1}{2E}
	\left[
	U
	\begin{pmatrix}
	m_1^2 & 0 & 0 & 0\\
	0 & m_2^2 & 0 & 0\\
	0 & 0 & m_3^2 & 0\\
	0 & 0 & 0 & m_4^2
	\end{pmatrix}
	U^\dagger
	+ \begin{pmatrix}
	a & 0 & 0 & 0\\
	0 & 0 & 0 & 0\\
	0 & 0 & 0 & 0\\
	0 & 0 & 0 & \eta a
	\end{pmatrix}
	\right]
	=
	\frac{1}{2E}
	V
	\begin{pmatrix}
	\widetilde m_1^2 & 0 & 0 & 0\\
	0 & \widetilde m_2^2 & 0 & 0\\
	0 & 0 & \widetilde m_3^2 & 0\\
	0 & 0 & 0 & \widetilde m_4^2
	\end{pmatrix}
	V^\dagger
	\;,
	\label{eq:fourHamiltonian}
	\end{align}
	where the sterile neutrino is supposed to participate in neither the neutral-current nor charged-current interaction, and $\eta\equiv N_n^{}/(2N_e^{})$, with $N_n^{}$ and $N_e^{}$ denoting the neutron and electron number densities, respectively. We assume a fixed matter composition, so $\eta$ remains constant as $a$ varies. Compared with the three-flavor case, the four-flavor framework has two essential differences. First, the real quartets are not uniquely fixed by the side lengths of the unitarity quadrangles and therefore cannot in general be eliminated in favor of $|V_{\alpha i}^{}|^2$. Consequently, the RGEs do not close on $\widetilde\Delta_{ij}^{}$ and $|V_{\alpha i}^{}|^2$ alone. Second, for $\eta\neq0$, the matter potential contains two independent flavor spurions, $P_e^{}=|e\rangle\langle e|$ and $P_s^{}=|s\rangle\langle s|$. Their rank-two combination $P_e^{}+\eta P_s^{}$ probes both the electron and sterile rows. These two features lead to RGE and invariant structures qualitatively different from those in the standard three-flavor framework.

	\subsection{The complete set of RGEs}
	\label{subsec:fourRGE}

	In fact, the four-flavor RGEs follow directly by repeating the derivation used in the three-flavor case. The complete set of RGEs for $\widetilde\Delta_{ij}^{}$ and $V_{\alpha i}^{}$ can be explicitly written as~\cite{Zeng:2022rxm}
	\begin{align}
	\frac{{\rm d}\widetilde\Delta_{ij}^{}}{{\rm d}a}
	&=
	|V_{ei}^{}|^2
	-|V_{ej}^{}|^2
	+\eta\left(
	|V_{si}^{}|^2
	-|V_{sj}^{}|^2
	\right)
	\;,
	\label{eq:fourMassRGE}\\
	\frac{{\rm d}V_{\alpha i}^{}}{{\rm d}a}
	&=
	\sum_{j\neq i}
	\frac{V_{\alpha j}^{}}{\widetilde\Delta_{ij}^{}}
	\left(
	V_{ej}^{*}V_{ei}^{}
	+\eta V_{sj}^{*}V_{si}^{}
	\right)
	\;,
	\label{eq:fourMixingRGE}
	\end{align}
	where $\alpha=e,\mu,\tau,s$ and $i,j=1,2,3,4$. To make direct contact with oscillation observables, we further rewrite Eq.~(\ref{eq:fourMixingRGE}) as the RGEs for $|V_{\alpha i}^{}|^2$, namely,
	\begin{equation}
	\frac{{\rm d}|V_{\alpha i}^{}|^2}{{\rm d}a}
	=
	2\sum_{j\neq i}
	\frac{
	\widetilde R_{e\alpha}^{ij}
	+\eta\widetilde R_{s\alpha}^{ij}}
	{\widetilde\Delta_{ij}^{}}
	\;,
	\label{eq:fourModuliRGE}
	\end{equation}
	where the real quartets $\widetilde R_{\rho\alpha}^{ij}$ are defined as in the three-flavor case. For a $4\times4$ unitary matrix, row orthogonality defines a quadrangle whose shape is not fixed by its side lengths. The real quartets therefore cannot in general be expressed purely by $|V_{\alpha i}^{}|^2$, and Eqs.~(\ref{eq:fourMassRGE}) and~(\ref{eq:fourModuliRGE}) do not form a closed system in $\widetilde\Delta_{ij}^{}$ and $|V_{\alpha i}^{}|^2$ alone. A closed set of RGEs then requires enlarging the variables to the complete set of bilinears $\widetilde{\cal A}_{\alpha\beta}^{i}\equiv V_{\alpha i}^{}V_{\beta i}^{*}$ with $\alpha,\beta=e,\mu,\tau,s$ and $i=1,2,3,4$. Consequently, we obtain
	\begin{align}
	\frac{{\rm d}\widetilde{\cal A}_{\alpha\beta}^{i}}{{\rm d}a}
	=
	\sum_{j\neq i}
	\frac{1}{\widetilde\Delta_{ij}^{}}
	\Bigl[
	\widetilde{\cal A}_{\alpha e}^{j}
	\widetilde{\cal A}_{e\beta}^{i}
	+
	\widetilde{\cal A}_{\alpha e}^{i}
	\widetilde{\cal A}_{e\beta}^{j}
	+\eta\left(
	\widetilde{\cal A}_{\alpha s}^{j}
	\widetilde{\cal A}_{s\beta}^{i}
	+
	\widetilde{\cal A}_{\alpha s}^{i}
	\widetilde{\cal A}_{s\beta}^{j}
	\right)
	\Bigr]
	\;,
	\label{eq:S4BilinearRGE}
	\end{align}
	where the RGEs for $|V_{\alpha i}^{}|^2$ are included because $\widetilde{\cal A}_{\alpha\alpha}^{i}
	=|V_{\alpha i}^{}|^2$. Eqs.~(\ref{eq:fourMassRGE}) and~(\ref{eq:S4BilinearRGE}) form a closed system of RGEs for the four-flavor framework. They retain their form under a simultaneous relabeling of the mass-eigenstate indices, reflecting an exact $S^{}_4$ permutation symmetry. 
	\subsection{The complete set of invariants}
	\label{subsec:fourInvariants}

	After the unitarity constraints are imposed and the flavor-rephasing redundancy is removed, Eqs.~(\ref{eq:fourMassRGE}) and~(\ref{eq:S4BilinearRGE}) span a twelve-dimensional physical space consisting of three independent $\widetilde\Delta_{ij}^{}$ and nine mixing parameters encoded in $\widetilde{\cal A}^{i}$, whereas evolution with $a$ selects one direction for fixed $\eta$. Hence a complete local set contains eleven functionally independent matter invariants. We now construct the complete set of polynomial matter invariants by combining the $S^{}_4$ permutation symmetry with flavor-rephasing invariance. 
	
	Similar to the three-flavor case, we first remove the identity shift from $\widetilde m_i^2$ by defining
	\begin{equation}
	\widetilde\ell_i^{}
	\equiv
	\widetilde m_i^2
	-
	\frac{1}{4}\sum_{j=1}^{4}\widetilde m_j^2
	\;.
	\label{eq:S4ell}
	\end{equation}
	By construction, $\sum_i\widetilde\ell_i^{}=0$, and $(\widetilde\ell_1^{},\widetilde\ell_2^{},\widetilde\ell_3^{},\widetilde\ell_4^{})^{\rm T}_{}$ therefore transforms in the three-dimensional standard representation $\boldsymbol 3$ of $S^{}_4$. It is easy to check that $\widetilde\ell_i^{}$ satisfies the following RGE
	\begin{equation}
	\frac{{\rm d}\widetilde\ell_i^{}}{{\rm d}a}
	=
	\widetilde{\cal A}_{ee}^{i}
	+\eta\widetilde{\cal A}_{ss}^{i}
	-
	\frac{1+\eta}{4}
	\;.
	\label{eq:S4ellRGE}
	\end{equation}

	The four components $\widetilde{\cal A}_{\alpha\beta}^{i}$ with $\alpha\neq\beta$ also satisfy $\sum_i\widetilde{\cal A}_{\alpha\beta}^{i}=0$ and transform in $\boldsymbol 3$. For $\alpha=\beta$, $\sum_i\widetilde{\cal A}_{\alpha\alpha}^{i}=1$, and the standard triplet representation is obtained by subtracting $1/4$ from each $\widetilde{\cal A}_{\alpha\alpha}^{i}$. Consequently, for each flavor pair $(\alpha,\beta)$, the tensor product of the $S^{}_4$ triplet $\widetilde\ell_i^{}$ with the corresponding triplet component of $\widetilde{\cal A}_{\alpha\beta}^{i}$ contains a unique $S^{}_4$ singlet, which is the lowest-order nontrivial singlet involving both quantities and is given by
	\begin{equation}
	\widetilde{\cal E}_{\alpha\beta}^{}
	\equiv
	\sum_{i=1}^{4}
	\widetilde\ell_i^{}
	\widetilde{\cal A}_{\alpha\beta}^{i}
	\;.
	\label{eq:S4FirstMoment}
	\end{equation}
	Note that the off-diagonal component $\widetilde{\cal E}_{\alpha\beta}^{}$ with $\alpha\neq\beta$ reads
	\begin{equation*}
	\widetilde{\cal E}_{\alpha\beta}^{}
	=
	\sum_{i=1}^{4}
	\widetilde m_i^2
	V_{\alpha i}^{}V_{\beta i}^{*}
	\;,
	\end{equation*}
	which is the direct four-flavor extension of $\widetilde Z_{\alpha\beta}^{}$ in Eq.~(\ref{eq:Zthree}).

	The RGE of $\widetilde{\cal E}$ can be derived from Eqs.~(\ref{eq:S4BilinearRGE}) and~(\ref{eq:S4ellRGE}). We obtain
	\begin{align}
	\frac{{\rm d}\widetilde{\cal E}}{{\rm d}a}
	={}&
	\sum_i
	\left(
	\widetilde{\cal A}_{ee}^{i}
	+\eta\widetilde{\cal A}_{ss}^{i}
	-
	\frac{1+\eta}{4}
	\right)
	\widetilde{\cal A}^{i}
	+\sum_i\sum_{j\neq i}
	\widetilde{\cal A}^{j}
	\left(P_e^{}+\eta P_s^{}\right)
	\widetilde{\cal A}^{i}
	\nonumber\\
	={}&
	\sum_{i,j}
	\widetilde{\cal A}^{i}
	\left(P_e^{}+\eta P_s^{}\right)
	\widetilde{\cal A}^{j}
	-
	\frac{1+\eta}{4}
	\sum_i\widetilde{\cal A}^{i}
	\nonumber\\
	={}&
	P_e^{}
	+\eta P_s^{}
	-
	\frac{1+\eta}{4}\boldsymbol 1\;,
	\label{eq:S4FirstMomentRGE}
	\end{align}
	where we have used $\widetilde{\cal A}^{i}(P_e^{}+\eta P_s^{})\widetilde{\cal A}^{i}=(\widetilde{\cal A}_{ee}^{i}+\eta\widetilde{\cal A}_{ss}^{i})\widetilde{\cal A}^{i}$ and $\sum_i\widetilde{\cal A}^{i}=\boldsymbol 1$. It is straightforward to verify that
	\begin{equation}
	P_e^{}+\eta P_s^{}-\frac{1+\eta}{4}\boldsymbol 1
	=
	\frac{1}{4}
	\begin{pmatrix}
	3-\eta & 0 & 0 & 0\\
	0 & -1-\eta & 0 & 0\\
	0 & 0 & -1-\eta & 0\\
	0 & 0 & 0 & 3\eta-1
	\end{pmatrix}
	\;,
	\label{eq:PeetaPsexplicit}
	\end{equation}
	which shows that the off-diagonal components of $\widetilde{\cal E}$ are independent of $a$, whereas its traceless diagonal sector evolves only along the direction fixed by the matter spurion. Exact matter invariants can then be constructed from rephasing-invariant combinations of $\widetilde{\cal E}$.

	First, the diagonal sector contains two independent matter invariants orthogonal to the direction selected by the matter spurion. A convenient basis is
	\begin{align}
	\widetilde D_{\mu\tau}^{}
	&\equiv
	\widetilde{\cal E}_{\mu\mu}^{}
	-
	\widetilde{\cal E}_{\tau\tau}^{}
	=
	\frac{1}{4}
	\sum_{i,j=1}^{4}
	\widetilde\Delta_{ij}^{}
	\left(
	|V_{\mu i}^{}|^2
	-
	|V_{\tau i}^{}|^2
	\right)
	\;,
	\nonumber\\
	\widetilde D_\eta^{}
	&\equiv
	\eta\widetilde{\cal E}_{ee}^{}
	+
	\frac{1-\eta}{2}
	\left(
	\widetilde{\cal E}_{\mu\mu}^{}
	+
	\widetilde{\cal E}_{\tau\tau}^{}
	\right)
	-
	\widetilde{\cal E}_{ss}^{}
	=
	\frac{1}{4}
	\sum_{i,j=1}^{4}
	\widetilde\Delta_{ij}^{}
	\left[
	\eta|V_{ei}^{}|^2
	+
	\frac{1-\eta}{2}
	\left(
	|V_{\mu i}^{}|^2
	+
	|V_{\tau i}^{}|^2
	\right)
	-
	|V_{si}^{}|^2
	\right]
	\;.
	\label{eq:fourDiagonalInvariants}
	\end{align}
	One can check that Eq.~(\ref{eq:S4FirstMomentRGE}) indeed leads to ${\rm d}\widetilde D_{\mu\tau}^{}/{\rm d}a=0$ and ${\rm d}\widetilde D_\eta^{}/{\rm d}a=0$ for fixed $\eta$. The invariant $\widetilde D_{\mu\tau}^{}$ already appears in the three-flavor framework, whereas $\widetilde D_\eta^{}$ arises from the introduction of the sterile neutrino $\nu_s^{}$.

	Next, invariants consisting of off-diagonal elements of $\widetilde{\cal E}$ can be obtained by arranging the flavor indices into closed rephasing-invariant cycles. For four flavors, the primitive cycles have lengths two, three, and four, which take the forms
	\begin{align}
	\widetilde{\cal Q}_{\alpha\beta}^{}
	&\equiv
	\widetilde{\cal E}_{\alpha\beta}^{}
	\widetilde{\cal E}_{\beta\alpha}^{}
	\;,
	\\
	\widetilde{\cal T}_{\alpha\beta\gamma}^{}
	&\equiv
	\widetilde{\cal E}_{\alpha\beta}^{}
	\widetilde{\cal E}_{\beta\gamma}^{}
	\widetilde{\cal E}_{\gamma\alpha}^{}
	\;,
	\\
	\widetilde{\cal W}_{\alpha\beta\gamma\delta}^{}
	&\equiv
	\widetilde{\cal E}_{\alpha\beta}^{}
	\widetilde{\cal E}_{\beta\gamma}^{}
	\widetilde{\cal E}_{\gamma\delta}^{}
	\widetilde{\cal E}_{\delta\alpha}^{}
	\;,
	\label{eq:S4CycleInvariants}
	\end{align}
	where the flavor labels are distinct. Eq.~(\ref{eq:S4FirstMomentRGE}) immediately gives
	\begin{equation}
	\frac{{\rm d}\widetilde{\cal Q}_{\alpha\beta}^{}}{{\rm d}a}
	=
	\frac{{\rm d}\widetilde{\cal T}_{\alpha\beta\gamma}^{}}{{\rm d}a}
	=
	\frac{{\rm d}\widetilde{\cal W}_{\alpha\beta\gamma\delta}^{}}{{\rm d}a}
	=0
	\;,
	\label{eq:S4CycleKernel}
	\end{equation}
	indicating that these rephasing-invariant cycles do not evolve with $a$. Now we identify the independent invariants from these cycles. 
	\begin{itemize}
		\item The six quantities $\widetilde{\cal Q}_{\alpha\beta}^{}$ with $\alpha < \beta$ correspond to the CP-even four-flavor sum rules $\widetilde {\cal Q}_{\alpha\beta}^{}={\cal Q}_{\alpha\beta}^{}$~\cite{Xing:2001bg,Zhang:2006yq}. They are precisely the counterparts of $\{\widetilde{\cal I}_2^{}, \widetilde{\cal I}_3^{}, \widetilde{\cal I}_4^{}\}$, or equivalently $\widetilde{B}^{}_{\alpha\beta}$ with $\alpha<\beta$, in the three-flavor framework. $\widetilde{\cal Q}_{\alpha\beta}^{}$ can be written explicitly in terms of $\widetilde\Delta_{ij}^{}$ and real quartets as
		\begin{align}
		\widetilde{\cal Q}_{\alpha\beta}^{}
		=
		\frac{1}{16}
		\sum_{i,j,k,l=1}^{4}
		\widetilde\Delta_{ij}^{}
		\widetilde\Delta_{kl}^{}
		\widetilde R_{\alpha\beta}^{ik}
		\;,
		\quad \alpha < \beta
		\;,
		\label{eq:S4CycleSpectralForms}
		\end{align}
		where the CP-odd quartet contributions cancel out pairwise under $i\leftrightarrow k$, leaving only $\widetilde R_{\alpha\beta}^{ik}$. The six $\widetilde{\cal Q}_{\alpha\beta}^{}$ therefore determine the magnitudes of all off-diagonal edges.
		\item $\widetilde{\cal T}_{\alpha\beta\gamma}^{}$ is invariant under cyclic permutations of $(\alpha,\beta,\gamma)$ and transforms into its complex conjugate under the exchange of any two flavor indices. Therefore four inequivalent triangle cycles remain, for which we
		choose the representatives $\{ \widetilde{\cal T}_{e\mu\tau}^{}, \widetilde{\cal T}_{e\mu s}^{}, \widetilde{\cal T}_{e\tau s}^{}, \widetilde{\cal T}_{\mu\tau s}^{}\}$. Their magnitudes satisfy
		\begin{equation}
		\left|\widetilde{\cal T}_{\alpha\beta\gamma}^{}\right|^2
		=
		\widetilde{\cal Q}_{\alpha\beta}^{}
		\widetilde{\cal Q}_{\beta\gamma}^{}
		\widetilde{\cal Q}_{\gamma\alpha}^{}
		\;,
		\label{eq:fourLoopRelation}
		\end{equation}
		so only the phases of $\widetilde{\cal T}_{\alpha\beta\gamma}^{}$ provide information beyond $\widetilde{\cal Q}_{\alpha\beta}^{}$. When all six $\widetilde{\cal Q}_{\alpha\beta}^{}$ are nonzero, they define the six nonvanishing edges of a tetrahedral flavor graph. Once three fundamental triangles, for example $\widetilde{\cal T}_{e\mu\tau}^{}$, $\widetilde{\cal T}_{e\mu s}^{}$, and $\widetilde{\cal T}_{e\tau s}^{}$, are chosen, the remaining triangle is uniquely determined by
		\begin{equation}
		\widetilde{\cal T}_{\mu\tau s}^{}
		=
		\frac{
		\widetilde{\cal T}_{e\mu\tau}^{}
		\widetilde{\cal T}_{e\tau s}^{}
		\widetilde{\cal T}_{e\mu s}^*}
		{
		\widetilde{\cal Q}_{e\mu}^{}
		\widetilde{\cal Q}_{e\tau}^{}
		\widetilde{\cal Q}_{es}^{}}
		\;.
		\label{eq:S4FourthTriangleRelation}
		\end{equation}
		Therefore, only three loop phases are independent.
		Specifically, we may take ${\rm Im}\,\widetilde{\cal T}_{e\mu\tau}^{}$, ${\rm Im}\,\widetilde{\cal T}_{e\mu s}^{}$, and ${\rm Im}\,\widetilde{\cal T}_{e\tau s}^{}$ as a local set of CP-odd invariants. Their explicit form is
		\begin{align}
		{\rm Im}\,\widetilde{\cal T}_{\alpha\beta\gamma}^{}
		=\frac{1}{64}
		\sum_{i,j,k,p,q,r=1}^{4}
		\widetilde\Delta_{ip}^{}
		\widetilde\Delta_{jq}^{}
		\widetilde\Delta_{kr}^{}
		{\rm Im}\left(
		V_{\alpha i}^{}V_{\beta i}^{*}
		V_{\beta j}^{}V_{\gamma j}^{*}
		V_{\gamma k}^{}V_{\alpha k}^{*}
		\right)
		\;,
		\label{eq:S4NaumovSpectralForm}
		\end{align}
		which equals its vacuum counterpart ${\rm Im}\,{\cal T}_{\alpha\beta\gamma}^{}$, giving rise to the extended Naumov relation in the four-flavor neutrino oscillations~\cite{Xing:2001bg}. Unlike the three-flavor relation, the four-flavor Naumov relation couples multiple CP-odd invariants and therefore does not admit a simple factorized form.
		\item Up to cyclic permutations and complex conjugation, there are three distinct quadrangle cycles $\widetilde{\cal W}_{\alpha\beta\gamma\delta}^{}$. Locally, they are fixed by $\widetilde{\cal T}_{\alpha\beta\gamma}^{}$ and $\widetilde{\cal Q}_{\alpha\beta}^{}$. For example,
		\begin{equation}
		\widetilde{\cal W}_{e\mu\tau s}^{}
		=
		\frac{
		\widetilde{\cal T}_{e\mu\tau}^{}
		\widetilde{\cal T}_{e\tau s}^{}}
		{\widetilde{\cal Q}_{e\tau}^{}}
		\;.
		\label{eq:S4CycleRelations}
		\end{equation}
		Thus, $\widetilde{\cal W}_{\alpha\beta\gamma\delta}^{}$ introduces no additional independent invariant except on the boundary where the relevant $\widetilde{\cal Q}_{\alpha\beta}^{}$ vanishes.
	\end{itemize}

	Taken together, the local invariant space can be parametrized by the following algebraically independent real quantities
	\begin{equation}
	\left\{
	\widetilde D_{\mu\tau}^{},
	\widetilde D_\eta^{},
	\widetilde{\cal Q}_{\alpha\beta}^{},
	{\rm Im}\,\widetilde{\cal T}_{e\mu\tau}^{},
	{\rm Im}\,\widetilde{\cal T}_{e\mu s}^{},
	{\rm Im}\,\widetilde{\cal T}_{e\tau s}^{}
	\right\}
	\;.
	\label{eq:fourCompleteSet}
	\end{equation}
	The two diagonal invariants $\{\widetilde D_{\mu\tau}^{},
	\widetilde D_\eta^{}\}$, six edge magnitudes $\widetilde{\cal Q}_{\alpha\beta}^{}$ with $\alpha < \beta$, and three independent triangle cycles $\widetilde{\cal T}_{\alpha\beta\gamma}^{}$ give $2+6+3=11$ invariants, matching the dimension count at the beginning of this subsection. This minimal local invariant set applies when all $\widetilde{\cal Q}_{\alpha\beta}^{}$ are nonzero and the signs of the selected ${\rm Re}\,\widetilde{\cal T}_{\alpha\beta\gamma}^{}$ are fixed. A complete global polynomial ring further requires the real and imaginary parts of all distinct $\widetilde{\cal T}_{\alpha\beta\gamma}^{}$ and $\widetilde{\cal W}_{\alpha\beta\gamma\delta}^{}$, together with algebraic relations such as the polynomial forms of Eqs.~(\ref{eq:S4FourthTriangleRelation}) and~(\ref{eq:S4CycleRelations}). These additional generators cover the boundary regions of parameter space where some edge magnitudes vanish and the associated phase coordinates are not well defined, without increasing the number of algebraically independent invariants.

	\subsection{No multiplicative invariant}
	\label{subsec:fourMultiplicative}

	Having characterized the invariant structure, we now follow the strategy used in the three-flavor framework to determine whether the four-flavor framework also admits compact multiplicative invariants. We first organize $\widetilde\Delta_{ij}^{}$ and $|V_{\alpha i}^{}|^2$ into $S^{}_4$ singlets
	\begin{equation}
	\widetilde{\cal V}_4^{}
	\equiv
	\prod_{i<j}\widetilde\Delta_{ij}^{}
	\;,
	\quad
	\widetilde\Pi_\alpha^{}
	\equiv
	\prod_{i=1}^{4}|V_{\alpha i}^{}|^2
	\;.
	\label{eq:fourS4Products}
	\end{equation}
	Then the most general monomial with a fixed $S^{}_4$ character is
	\begin{equation}
	\widetilde{\cal M}_4^{}
	=
	\widetilde{\cal V}_4^p
	\widetilde\Pi_e^q
	\widetilde\Pi_\mu^r
	\widetilde\Pi_\tau^s
	\widetilde\Pi_s^t
	\;,
	\quad
	p,q,r,s,t\in\mathbb N
	\;,
	\label{eq:fourMonomial}
	\end{equation}
	which transforms with character $(-1)^p$. At the pole $\widetilde\Delta_{ij}^{}=0$, the residue of ${\rm d}\ln\widetilde{\cal M}_4^{}/{\rm d}a$ is
	\begin{align}
	\left.
	{\rm Res}\left(
	\frac{{\rm d}\ln\widetilde{\cal M}_4^{}}{{\rm d}a}
	\right)
	\right|_{\widetilde\Delta_{ij}^{}=0}
	={}&
	p\left[
	|V_{ei}^{}|^2
	-|V_{ej}^{}|^2
	+\eta\left(
	|V_{si}^{}|^2
	-|V_{sj}^{}|^2
	\right)
	\right]
	\nonumber\\
	&+2\sum_{\alpha=e,\mu,\tau,s}
	n_\alpha^{}
	\left(
	\widetilde R_{e\alpha}^{ij}
	+\eta\widetilde R_{s\alpha}^{ij}
	\right)
	\left(
	\frac{1}{|V_{\alpha i}^{}|^2}
	-\frac{1}{|V_{\alpha j}^{}|^2}
	\right)
	\;,
	\label{eq:fourPoleResidue}
	\end{align}
	where $n_e^{}=q$, $n_\mu^{}=r$, $n_\tau^{}=s$, and $n_s^{}=t$. The $\mu$- and $\tau$-row contributions contain independent real quartets, so cancellation for generic mixing first requires $r=s=0$, consistent with the three-flavor case. The remaining $e$- and $s$-row terms contain $\widetilde R_{es}^{ij}$ with coefficient
	\begin{equation}
	2\left[
	\eta q
	\left(
	\frac{1}{|V_{ei}^{}|^2}
	-\frac{1}{|V_{ej}^{}|^2}
	\right)
	+t
	\left(
	\frac{1}{|V_{si}^{}|^2}
	-\frac{1}{|V_{sj}^{}|^2}
	\right)
	\right]
	\;.
	\label{eq:fourMixedCoefficient}
	\end{equation}
	For $\eta\neq0$, this coefficient vanishes identically only if $q=t=0$. The remaining $1/\widetilde\Delta_{ij}^{}$ residue then requires $p=0$. Therefore
	\begin{equation}
	p=q=r=s=t=0
	\;,
	\label{eq:fourMonomialNoGo}
	\end{equation}
	showing that no nontrivial exact monomial constructed only from $\widetilde\Delta_{ij}^{}$ and $|V_{\alpha i}^{}|^2$ exists in the physical rank-two case. The absence of a monomial invariant in the $e$--$s$ sector can be understood from the following Gram determinant of the matter spurion projected onto each $(i,j)$ mass subspace
	\begin{align}
	\left|
	\begin{matrix}
	[V^\dagger(P_e^{}+\eta P_s^{})V]_{ii}^{}
	&
	[V^\dagger(P_e^{}+\eta P_s^{})V]_{ij}^{}
	\\
	[V^\dagger(P_e^{}+\eta P_s^{})V]_{ji}^{}
	&
	[V^\dagger(P_e^{}+\eta P_s^{})V]_{jj}^{}
	\end{matrix}
	\right|
	=\eta
	\left|
	V_{ei}^{}V_{sj}^{}
	-V_{ej}^{}V_{si}^{}
	\right|^2
	\;,
	\label{eq:fourGramObstruction}
	\end{align}
	which is generically nonzero for $\eta\neq0$. As a consequence, the electron and sterile projections in each $(i,j)$ mass subspace are linearly independent. This linear independence induces the different cofactors in Eq.~(\ref{eq:fourMixedCoefficient}), which cannot cancel out and thereby preclude any nontrivial monomial invariant. This conclusion remains unchanged if the real and imaginary parts of the quartets are included in Eq.~(\ref{eq:fourMonomial}), since their logarithmic derivatives develop independent poles at generic zeros of their real or imaginary parts, forcing all corresponding exponents to vanish. In the $\eta=0$ limit, however, Eq.~(\ref{eq:fourGramObstruction}) vanishes and the four-flavor electron-row product
	\begin{equation}
	\widetilde I_{e,4}^{}
	\equiv
	\prod_{i=1}^{4}|V_{ei}^{}|^2
	\prod_{i<j}\widetilde\Delta_{ij}^{2}
	\;,
	\label{eq:fourRankOneInvariant}
	\end{equation}
	becomes an exact matter invariant.

	\section{Summary}
	\label{sec:summary}

	In this paper, we formulate the search for exact matter invariants as the determination of first integrals of the RGEs for neutrino oscillations in matter. By combining the exact permutation covariance under relabeling of the mass eigenstates with flavor-rephasing invariance, we directly decipher the invariant structure and its multiplicative sector from the RGEs.

	In the three-flavor framework, after removing the unobservable trace from the effective mass-squared eigenvalues and centering the moduli of the mixing matrix, we organize $\widetilde\Delta_{ij}^{}$ and $|V_{\alpha i}^{}|^2$ into $S^{}_3$ singlets which are invariant under the simultaneous relabeling of the neutrino mass eigenstates. Their closed RGEs yield five algebraically independent first integrals, which constitute a complete set of matter invariants. After physical identification, these first integrals can be identified with the invariants constructed from the matter-independent $H_{\alpha\beta}^{}$ for $(\alpha,\beta)\neq(e,e)$, indicating that the RGE and Hamiltonian constructions give the same invariant ring.

	We then investigate independent multiplicative invariants. We organize $\widetilde\Delta_{ij}^{}$, $|V_{\alpha i}^{}|^2$, and $\widetilde J$ into the factorized $S^{}_3$ singlets $\widetilde{\cal V}_3^{}$, $\widetilde\Pi_\alpha^{}$, and $\widetilde J$. The rank-one matter spurion $P_e^{}$ probes only the electron row and leaves the orthogonal $\mu$--$\tau$ subspace invariant, which ensures that the RGEs for $\widetilde\Delta_{ij}^{}$ and $|V_{ei}^{}|^2$ form a closed subsystem. Consequently, $\widetilde{\cal V}_3^{}$, $\widetilde\Pi_e^{}$, and $\widetilde J$ are governed by the same pole cofactor with fixed relative coefficients. Canceling their pole terms leads to only two independent monomial invariants, $\widetilde I_{\rm N}^{}\equiv\widetilde J\widetilde\Delta_{12}^{}\widetilde\Delta_{23}^{}\widetilde\Delta_{31}^{}$ and $\widetilde I_e^{}\equiv|V_{e1}^{}|^2|V_{e2}^{}|^2|V_{e3}^{}|^2\widetilde\Delta_{12}^2\widetilde\Delta_{23}^2\widetilde\Delta_{31}^2$.
	The former gives the Naumov relation, while the Toshev relation follows from $\widetilde I_{\rm T}^{}=\widetilde I_{\rm N}^{}/(\widetilde I_e^{})^{1/2}$, leaving no additional independent multiplicative relation.

	We further investigate the four-flavor extension. Unlike the three-flavor case, the RGEs in the four-flavor framework do not close on $\widetilde\Delta_{ij}^{}$ and $|V_{\alpha i}^{}|^2$, so the mixing parameters must be enlarged to the complete bilinears $\widetilde{\cal A}_{\alpha\beta}^{i}\equiv V_{\alpha i}^{}V_{\beta i}^{*}$, from which the $S^{}_4$ singlets are constructed. Combining their closed RGEs with flavor-rephasing invariance gives a complete set of eleven algebraically independent matter invariants. As for the multiplicative sector, the matter spurion has a rank-two electron--sterile structure, whose electron and sterile projections in each mass subspace are generically linearly independent. The different factors therefore do not share a common pole cofactor, preventing the required cancellation and excluding any nontrivial multiplicative monomial invariant for generic mixing at fixed $\eta\neq0$.

	\section*{Acknowledgments}
	
	XW was funded by the European Union, NextGenerationEU, National Recovery and Resilience Plan (mission 4, component 2)
	under the project \textit{MODIPAC: Modular Invariance in Particle Physics and Cosmology} (CUP C93C24004940006). SZ was supported in part by the National Natural Science Foundation of China under grant No.~12475113 and No.~12535007, by the CAS Project for Young Scientists in Basic Research (YSBR-099), and by the Scientific and Technological Innovation Program of IHEP under grant No.~E55457U2.

    \bibliographystyle{JHEP}
	\bibliography{Ref}

\end{document}